\documentclass[conference]{IEEEtran}
\IEEEoverridecommandlockouts
\usepackage{cite}
\usepackage{amsmath,amssymb,amsfonts}
\usepackage{algorithmic}
\usepackage{graphicx}
\usepackage{textcomp}
\usepackage{xcolor}

\ifCLASSOPTIONcompsoc
  \usepackage[caption=false,font=normalsize,labelfont=sf,textfont=sf]{subfig}
\else
  \usepackage[caption=false,font=footnotesize]{subfig}
\fi

\def\BibTeX{{\rm B\kern-.05em{\sc i\kern-.025em b}\kern-.08em
    T\kern-.1667em\lower.7ex\hbox{E}\kern-.125emX}}
\begin{document}

\title{Deep Learning-Based Constellation Optimization for Physical Network Coding in \\Two-Way Relay Networks}

\author{
\IEEEauthorblockN{
Toshiki Matsumine\IEEEauthorrefmark{1}\IEEEauthorrefmark{2},
Toshiaki Koike-Akino\IEEEauthorrefmark{1}, and
Ye Wang\IEEEauthorrefmark{1}}
\IEEEauthorblockA{
\IEEEauthorrefmark{1}
Mitsubishi Electric Research Laboratories (MERL), 201 Broadway, Cambridge, MA 02139, USA\\
\IEEEauthorrefmark{2}
Department of Electrical and Computer Engineering, Yokohama National University, Kanagawa, 240-8501, Japan\\
Email: matsumine-toshiki-tk@ynu.jp, \{koike, yewang\}@merl.com}
}

\maketitle

\begin{abstract}
This paper studies a new application of deep learning (DL) for optimizing constellations in two-way relaying with physical-layer network coding (PNC),
where deep neural network (DNN)-based modulation and demodulation are employed at each terminal and relay node.
We train DNNs such that the cross entropy loss is directly minimized, and thus it maximizes the likelihood,
rather than considering the Euclidean distance of the constellations.
The proposed scheme can be extended to higher level constellations with slight modification of the DNN structure.
Simulation results demonstrate a significant performance gain in terms of the achievable sum rate over conventional relaying schemes.
Furthermore, since our DNN demodulator directly outputs bit-wise probabilities, it is straightforward to concatenate with soft-decision channel decoding.
\end{abstract}


\section{Introduction}
Physical-layer network coding (PNC) is a promising approach for increasing a throughput in wireless relay networks with two-way,
or more generally, multi-way communication flows.
Several protocols for two-way wireless relaying schemes have been proposed and analyzed in the literature \cite{laneman2004cooperative,popovski2006anti,popovski2007physical,kim2008performance,rankov2007spectral,louie2010practical,wilson2010joint,zhang2009channel,koike2009denoising,liew2013physical}.
These approaches can be largely classified depending on the number of steps it takes for user terminals to exchange their packets via a relay in the middle.
For example, amplify-and-forward (AF) is a well-known approach for 2-step protocols
where two terminals send their packets to a relay in the first step, and the relay simply broadcasts the amplified version of the received signal.
On the other hand, in 3-step decode-and-forward schemes, both terminals send their packets in different steps to avoid interference,
and then the relay broadcasts the signal after decoding the packets from each terminal.
In this paper, we investigate two-way relaying with 2-step protocols,
which potentially achieves a higher throughput than other protocols due to its time efficiency.

Optimized constellations for two-way relaying with a 2-step protocol based on denoise-and-forward (DNF) has been investigated in \cite{koike2009optimized}.
It is shown that optimized nonlinear network coding in terms of the Euclidean distance property offers significant improvement of system throughput compared to conventional network coding based on exclusive-or (XOR).
However, the major challenge of the strategy therein is
that the Euclidean distance of the constellation is not necessarily the most important performance measure in practical systems,
where binary soft-decision channel decoding is employed.
More specifically, for such systems, generalized mutual information (GMI) may be the appropriate measure that should be maximized.
Furthermore, since the network coding in \cite{koike2009optimized} is highly optimized for various factors,
such as channel coefficients and the combination of constellations transmitted from two terminals,
it requires a large look-up table to implement and thus it may be infeasible as the constellation size increases.

Inspired by the recent advancements of deep learning (DL) techniques, the application of DL has been widely studied in communication systems
\cite{kim2018communication, farsad2018neural, farsad2018sliding, farsad2018deep, gruber2017deep, liang2018iterative, o2017introduction, dorner2018deep, karanov2018end, aoudia2018end, ye2018channel}.
For example, DNN-based signal detection \cite{farsad2018neural, farsad2018sliding},
joint source-channel coding \cite{farsad2018deep}, channel decoding \cite{gruber2017deep, liang2018iterative} and DNN-based autoencoder \cite{o2017introduction, dorner2018deep, karanov2018end, aoudia2018end} were investigated.
More recently, end-to-end learning of communication systems based on generative adversarial networks was introduced in \cite{ye2018channel}.

In this paper, we investigate a new application of DL to   optimizing constellations for two-way relaying capable of nonlinear PNC.
We consider two-way wireless relaying with 2-step protocols, where all modulation/demodulation at terminals and a relay are performed by DNNs.
Unlike the conventional strategy for constellation optimization based on the Euclidean distance in the signal space,
our objective of training is to minimize the cross entropy, and thus it directly maximizes the GMI,
which is an important metric for a system employing soft-decision forward error correction.
Our proposed DNN-based approach can be easily extended to the higher order constellations,
and furthermore, the output of the DNN demodulator can be immediately fed into the soft-decision channel decoder at the receiver,
which is attractive in practice because we do not require an additional converter to generate log-likelihood ratios.

\begin{figure*}[t]
\centering
\includegraphics[width=0.75\hsize]{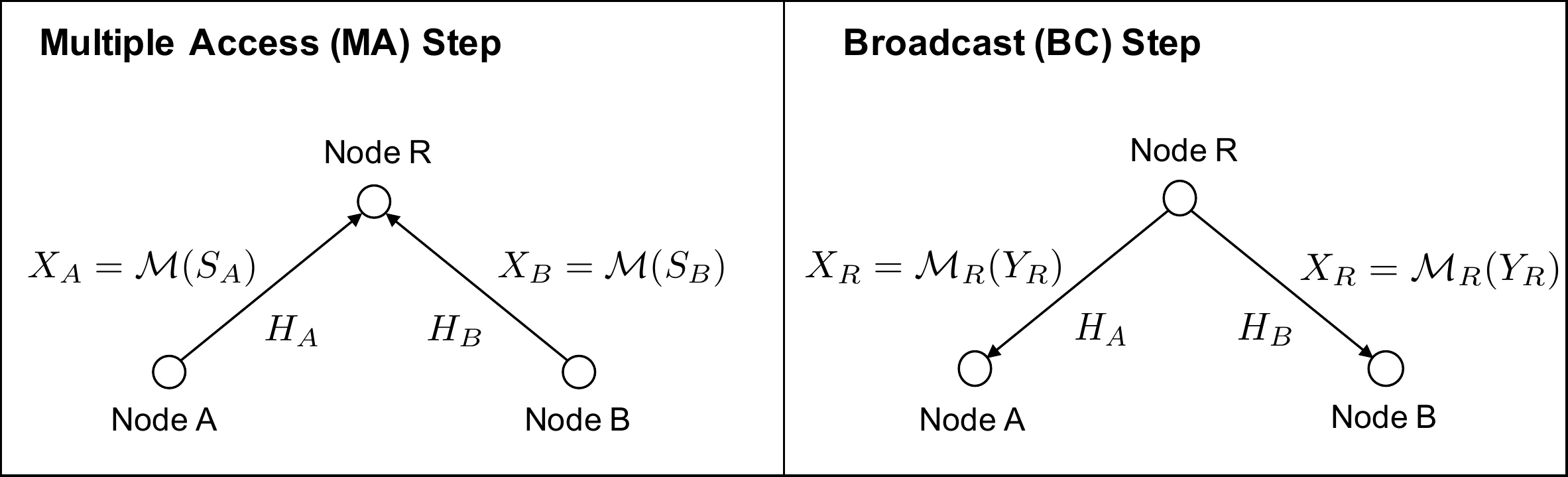}
\caption{2-step two-way relay wireless communication systems with PNC.}
\label{fig:system}
\end{figure*}

The contributions of this paper are summarized as follows:
\begin{itemize}
\item We propose a new nonlinear network coding approach based on DNN for two-way relay networks.
\item We design modulation and demodulation functions via DL such that the cross entropy loss is directly minimized in an end-to-end manner.
\item We extend our DNN design to high order modulation.
\item We demonstrate from simulation results that our DL-based approach significantly improves the achievable sum rate over the conventional counterparts.
\end{itemize}

The rest of this paper is organized as follows:
The general system model of two-way relay networks with 2-step protocols
is introduced in Section~\ref{sec:system}.
Section~\ref{sec:learning} describes the proposed relay model based on
the DNN modulation/demodulation and their learning process.
In Section~\ref{sec:sim},
we evaluate the performance of the proposed scheme
in terms of the achievable sum rates,
where better performance than conventional relaying is observed.
Finally, concluding remarks are given in Section~\ref{sec:con}.

\section{Two-Way Relay Systems with 2-Step Protocols}
\label{sec:system}
Fig.~\ref{fig:system} shows the system model for 2-step two-way wireless relaying, where terminal A has packets for terminal B, and vice versa.
The relay node R performs PNC to assist the data exchange, but it is neither a traffic source nor a sink for simplicity. 

\subsection{Multiple Access (MA) Step}
Letting $\mathcal{M}$ be a signal mapping function, the transmitting signals at terminals A and B are given by
$X_A=\mathcal{M}(S_A)$ and $X_B=\mathcal{M}(S_B)$, respectively, where $S_A$ and $S_B$ are binary source data per symbol at each terminal.
We assume that the mapping functions for terminals A and B are identical, and also,
neither terminal can receive anything from the other terminal during the MA stage due to the half–duplex constraint.
Then the received signal at the relay node R is expressed as $Y_R=H_AX_A+H_BX_B+Z_R$,
where $H_A$ and $H_B$ are complex channel coefficients from each terminal, and $Z_R$ is additive white Gaussian noise (AWGN).

\subsection{Broadcast (BC) Step}
The received signal at relay R, $Y_R$, is processed by the relay function $\mathcal{M}_R$, and then broadcasted to terminals A and B.
This function depends on a specific protocol, for example, in the case of the conventional AF relaying, this function is just a linear scaling function.
We consider nonlinear PNC functions for $\mathcal{M}_R$, designed by DL.  We denote the transmitting signal from the relay R as $X_R=\mathcal{M}_R(Y_R)$.
The received signals at terminals A and B are then given by $Y_A=H_AX_R+Z_A$ and $Y_B=H_BX_R+Z_B$, respectively,
where $Z_A$ and $Z_B$ are AWGNs.
For simplicity, we assume the reciprocal channel for both steps.
Finally, letting $\mathcal{D}$ be the demodulator function, terminals A and B detect their desired data by the demodulator as
$\hat{S}_A=\mathcal{D}(Y_B)$ and $\hat{S}_B=\mathcal{D}(Y_A)$, respectively. Note that the node A can exploit what he sent ($X_A$) to demodulate $X_B$, and similarly for node B.

\begin{figure*}[t]
\centering
\includegraphics[width=0.6\hsize]{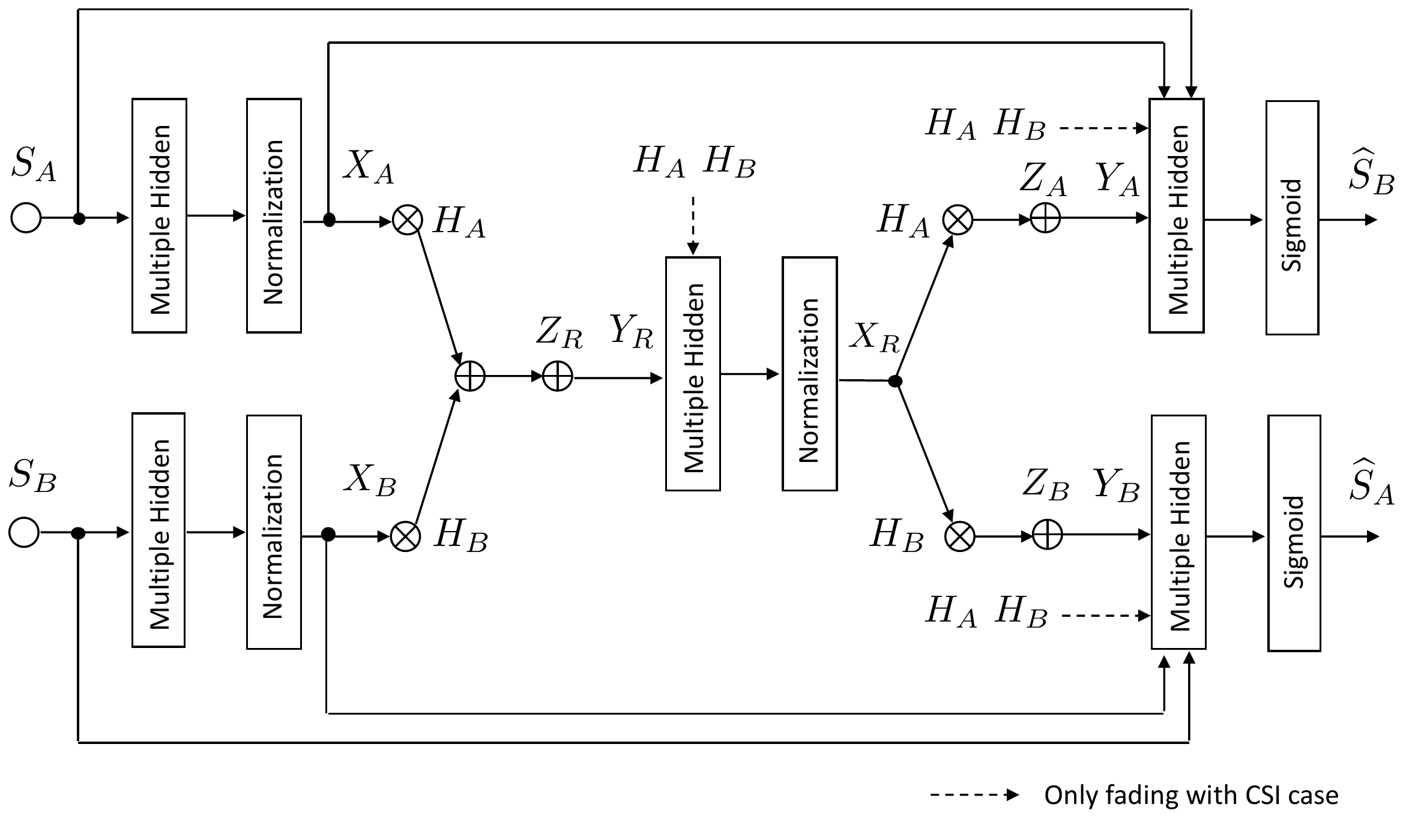}
\caption{Block diagram of DNN-based 2-step two-way relay systems.}
\label{fig:diagram}
\end{figure*}

\subsection{Main Challenges in Two-Way Relay Protocols}
The major challenge of the design of PNC for 2-step two-way relaying lies in
how to deal with the interference between the signals from two terminals in the MA step.
In the conventional AF relaying scheme, the relay just amplifies the received signal and then broadcasts it to the terminals A and B in the BC step.
This AF relaying can cause noise enhancement.
Furthermore, transmitting the amplified version of the summation of $X_A$ and $X_B$ (plus noise)  to each terminal may be inefficient,
since terminals A and B already have their own information $X_A$ and $X_B$, respectively.

To cope with these issues, DNF was proposed in \cite{popovski2006bi},
where the relay first detects two signals from the terminals A and B by maximum-likelihood (ML) detection,
and then maps into a discrete signal constellation.
In order to show the fundamental concept of DNF, let us assume the scenario where each terminal sends BPSK signals to the relay in the MA step,
i.e., $X_A, X_B \in \{-1, 1\}$.
For simplicity of explanation, we assume $H_A=H_B=1$ and a noiseless channel.
In this case, the possible received signal at the relay is $Y_R=\{-2, 0, 2\}$.
If the relay receives $Y_R=0$, there exists residual ambiguity and the relay cannot detect $X_A$ and $X_B$.
However, we may use the following denoising map in the BC step: $X_R=1$ for $Y_R=\{-2, 2\}$ and $X_R=-1$ otherwise.
When the terminal A has sent $X_A=1$ at the MA stage,  the relay will transmit $X_R=1$ if the terminal B also has sent $X_B=1$, and $X_R=1$ is transmitted if the terminal B has sent $X_B=-1$.
This is how the terminal A can successfully decide the information of the terminal B, and vice versa in DNF.

The design strategy of the denoising map for DNF scheme that optimizes the Euclidean property has been investigated in \cite{koike2009optimized}.
It is shown in \cite{koike2009optimized} that the optimization of signal constellations for PNC schemes in two-way relay networks
offers the significant performance gain over conventional XOR-based network coding in terms of the end-to-end throughput.
However, the major challenge of this strategy includes efficient implementation when the constellation size increases,
since it requires a large number of denoising mapping patterns.
Furthermore, the Euclidean distance is not necessarily of primary interest when channel codes are used.
For this reason, we propose the DNN-based modulation/demodulation scheme,
that directly minimizes the bit-wise cross entropy, or equivalently, maximizes the bit likelihood.
Furthermore, the proposed DNN-based scheme may be suited for efficient hardware implementation and scalable to higher level modulation.

\section{Deep Learning for PNC in Two-Way Relay}
\label{sec:learning}

In this section, we introduce the DNN-based two-way relay system and describe the learning process for DNN-based modulator/demodulator/denoiser optimization.
Fig.~\ref{fig:diagram} depicts the proposed system, where we have three DNNs  that are jointly trained.

\subsection{DNN Architecture}

\subsubsection{Modulator at Terminals}
Let $\mathcal{M}: \{0,1\}^k \rightarrow \mathbb{R}^n$ denote the DNN modulator function at the terminals,
which maps $k$-binary bits $S \in \{0, 1\}^k$ into the constellation in the $n$-dimensional real space $X \in \mathbb{R}^n$,
i.e., $X = \mathcal{M}( S )$.
We use the multi-layer perceptron (MLP) with multiple hidden layers, each of which has a number of nonlinear activation nodes.
At the last layer, the signal power of $X$ is normalized with a batch normalization layer.

\subsubsection{PNC Modulator at Relay}
Let $\mathcal{M}_R: \mathbb{R}^n \rightarrow \mathbb{R}^n$ denote the DNN modulator function at the relay,
which maps the real-valued $n$-dimensional vector $Y_R \in \mathbb{R}^n$
into the constellation in the $n$-dimensional real space $X_R \in \mathbb{R}^n$,
i.e., $X_R = \mathcal{M}_R( Y_R )$.
Note that the DNN modulator based on MLP can perform as a nonlinear denoising map, capable of PNC function.
Analogously to $\mathcal{M}$, we control the signal power with batch normalization at the output of the DNN modulator.

\subsubsection{Demodulator at Terminals}
A DNN demodulator function at terminals, denoted as $\mathcal{D}:  \mathbb{R}^n \rightarrow \mathbb{R}^k $,
maps the signal in the $n$-dimensional real space $Y \in \mathbb{R}^n$
into the $k$-dimensional probability vector $\widehat{S} \in \mathbb{R}^k$.
Since at each terminal, their own information are available, these information are used as an input to the DNN by appending as $\widehat{S}_B = \mathcal{D}( S_A, X_A, Y_A )$ and $\widehat{S}_A = \mathcal{D}( S_B, X_B, Y_B )$ at terminals A and B, respectively.
The outputs of the hidden layers is scaled into the range of $[0, 1]$ at the last sigmoid layer, which is interpreted as a likelihood.

Throughout this work, we use feedforward DNNs with 2 hidden layers with 1000 nodes, except for the last layer of the transmitter and receiver, which
are linear and sigmoid, respectively.
For simplicity, we assume $n=2$ in this work to represent in-phase and quadrature signaling
as an extension to higher-dimensional modulations is straightforward.

\subsection{End-to-End Learning}
The objective of the training is to minimize the cross entropy loss
between the binary data transmitted from each terminal and the decoded information at each terminal after relaying.
In order to do this, we jointly train three DNNs, i.e., the modulator $\mathcal{M}$ at the terminal, the PNC modulator $\mathcal{M}_R$ at the relay, and the demodulator $\mathcal{D}$ at the terminal.

With stochastic gradient descent, we minimize the following binary cross entropy loss:
\begin{align}
\mathcal{L}_\mathrm{loss}
& 
= \mathbb{E}_{b,l}\bigl[- (1-b) \log_2 (l) - b \log_2 (1-l) \bigr] \notag\\
&= \mathbb{E}_{b,L}\bigl[ \log_2\bigl( 1 + \exp(-(-1)^b L)\bigr) \bigr],
\label{eq:bce}
\end{align}
where the expectation $\mathbb{E}[\cdot]$ is taken over all transmitting bit $b \in \{0, 1\}$  and demodulator output $l$ across the training data. Here, we let $L$ denote the demodulator output prior to the sigmoid function as $l = 1/(1+\exp(-L))$. 
As shown in the following subsection, minimizing \eqref{eq:bce} is equivalent to maximizing the achievable rate in terms of GMI.

\subsection{Relationship Between Cross Entropy and GMI}
The cross entropy given in \eqref{eq:bce} is minimized by increasing the log-probability corresponding to the information bit,
which is equivalent to improving the bit log-likelihood ratio (LLR).
The normalized GMI in data communications is given as
\begin{align}
\mathsf{GMI} =
1 - \mathbb{E}_{b,L} \left[ \log_2 \left( 1 + \exp( -(-1)^b L ) \right) \right],
\label{eq:gmi}
\end{align}
where $L=\Pr(b=0)/\Pr(b=1)$ is the LLR value corresponding to the information bit $b$.
From \eqref{eq:bce} and \eqref{eq:gmi}, we can see that the GMI is a function of cross entropy over empirical ensemble: $\mathsf{GMI} = 1 - \mathcal{L}_\mathrm{loss}$.
Therefore we attempt to minimize the bit-wise cross entropy loss for maximizing GMI so that the achievable sum rates in two-way relaying systems can be maximized.

\subsection{Learning Fading from CSI}
When the channel state information (CSI) is available at each terminal and relay,
we can incorporate the channel coefficients $H_A$ and $H_B$ as an input to the modulators.
For example, when the channel coefficients $H_A$ and $H_B$ are available at the relay,
we can expand the input dimension of the DNN modulator at the relay as
\begin{align}
X_R = \mathcal{M}_R( Y_R, H_A, H_B).
\end{align}
Rather than such a simple concatenation of side information, it was found that further expansion of the input dimension with various different patterns of the data set leads to better performance. For example, the input dimensions of the DNN modulator at the relay may be extended by 
using the original input multiplied by channel coefficients $H_A, H_B$ as
\begin{align}
 X_R = \mathcal{M}_R( Y_R, H_A, H_AY_R, H_B, H_BY_R).
\label{eq:input_relay}
\end{align}
We note that increasing input dimension may decelerate the learning convergence of the DNN, while improves the performance.
In this paper, in addition to \eqref{eq:input_relay}, we use ($H^*_AY_R, H^*_BY_R, H^{-1}_AY_R, H^{-1}_BY_R$)
as the input to the DNN modulator at the relay when perfect CSI is available.

Similarly, we may use channel coefficients as inputs to the DNN demodulator at terminals A and B.
We assume that the CSI is used only for demodulation (not for modulation) at the terminals when it is available.

\section{Simulation Results}
\label{sec:sim}
In this section, we evaluate the performance of the proposed DNN-based PNC scheme.
We compare the performance of the proposed scheme with that of the conventional AF relaying
in both the AWGN and frequency-flat block Rayleigh fading channel.
We vary an average SNR $\mathbb{E}[|H_A|^2+|H_B|^2]/2\sigma^2$ for the performance evaluation,
where $\sigma^2$ is the variance of the Gaussian noise.
The channel power ratio between $\mathbb{E}[|H_A|^2]$ and $\mathbb{E}[|H_B|^2]$ is assumed to be $0$\,dB.

We use chainer \cite{chainer_learningsys2015} for the DNN implementation.
We set a mini-batch size as $128$ and generate $128 \times 1000$ pseudo random bits as a training data.
The epoch size is chosen as $30$ in all cases.
In each epoch the gradient of the loss function is calculated over the whole training data using Adam \cite{kingma2014adam} with learning rate $0.001$.
The DNN modulator/demodulator are trained using randomly generated channel coefficients and Gaussian noises, where
the same channel coefficients are used over a mini-batch, while Gaussian noise varies from symbol to symbol.
Note that 
our system is trained through randomly varying channel coefficients so that the designed PNC works even for practical channels.

As mentioned earlier, bit/block error rate may not be the appropriate performance measure
from the fact that the practical wireless system typically uses soft-decision forward error correction such as low-density parity-check codes.
For this reason, we focus on the achievable sum rate for the performance evaluation as
\begin{align}
2 K (1-\mathbb{E}[\mathcal{L}_\mathrm{loss}]) \ \text{[bps/Hz]},
\label{eq:thp}
\end{align}
where $2$ comes from the number of terminals and $K$ is the number of bits per symbol.

The training SNR is an important parameter when we train the proposed DNN-based system.
In the following, we train the DNN at every $5$\,dB, and then plot the best point among them for a given channel SNR.
In practice, this can be regarded as the performance with the adaptive selection of mapping/demapping functions depending on the estimated channel SNR.

\begin{figure}[t]
\centering
\includegraphics[width=0.9\hsize]{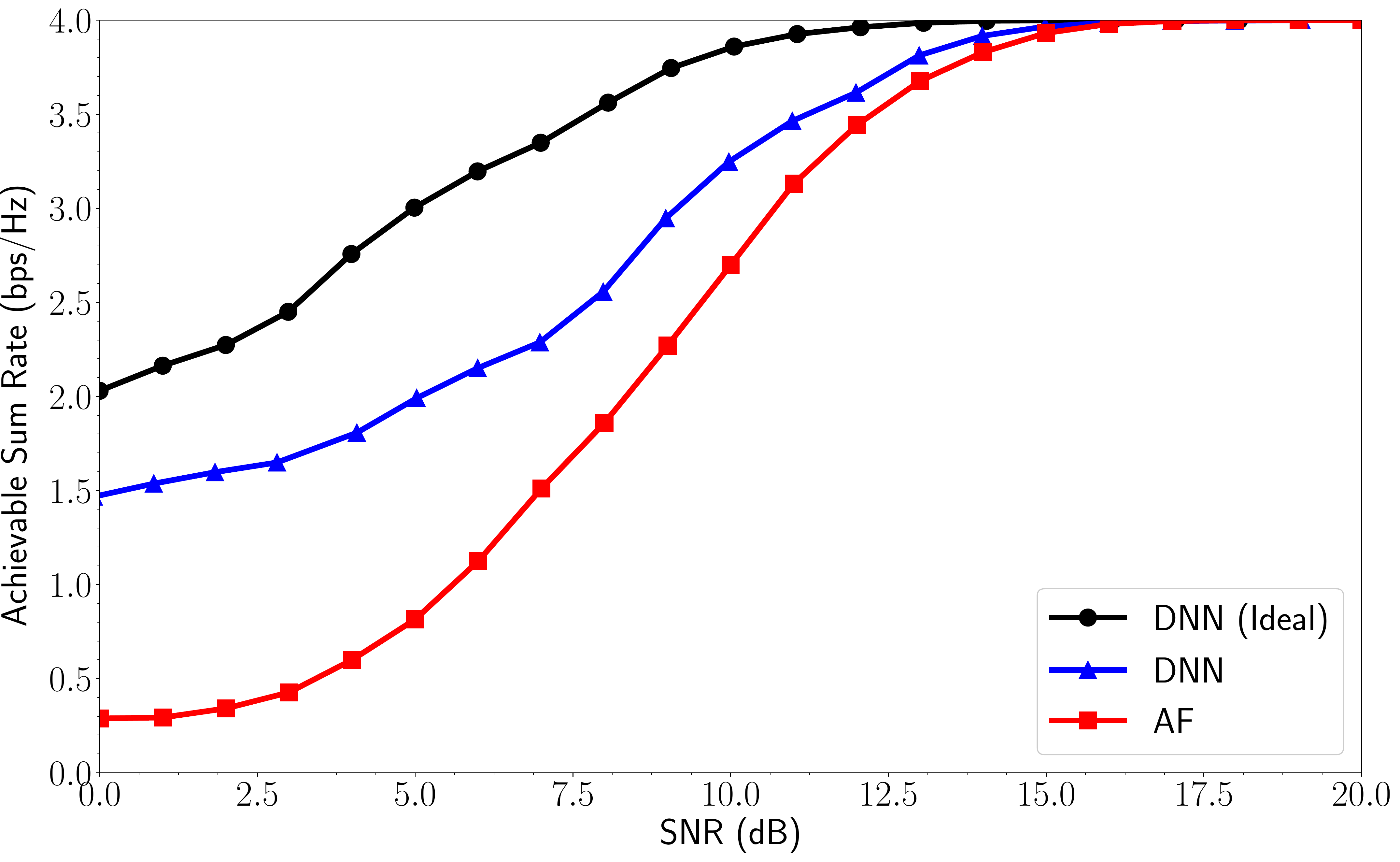}
\caption{Achievable sum rates of the proposed DNN relaying and the conventional AF relaying over AWGN channel with 4-QAM.}
\label{fig:qpsk_awgn}
\end{figure}
\begin{figure}[t]
 \centering
 \subfloat[Training SNR: $0$~dB]{ \includegraphics[width=0.45\hsize]{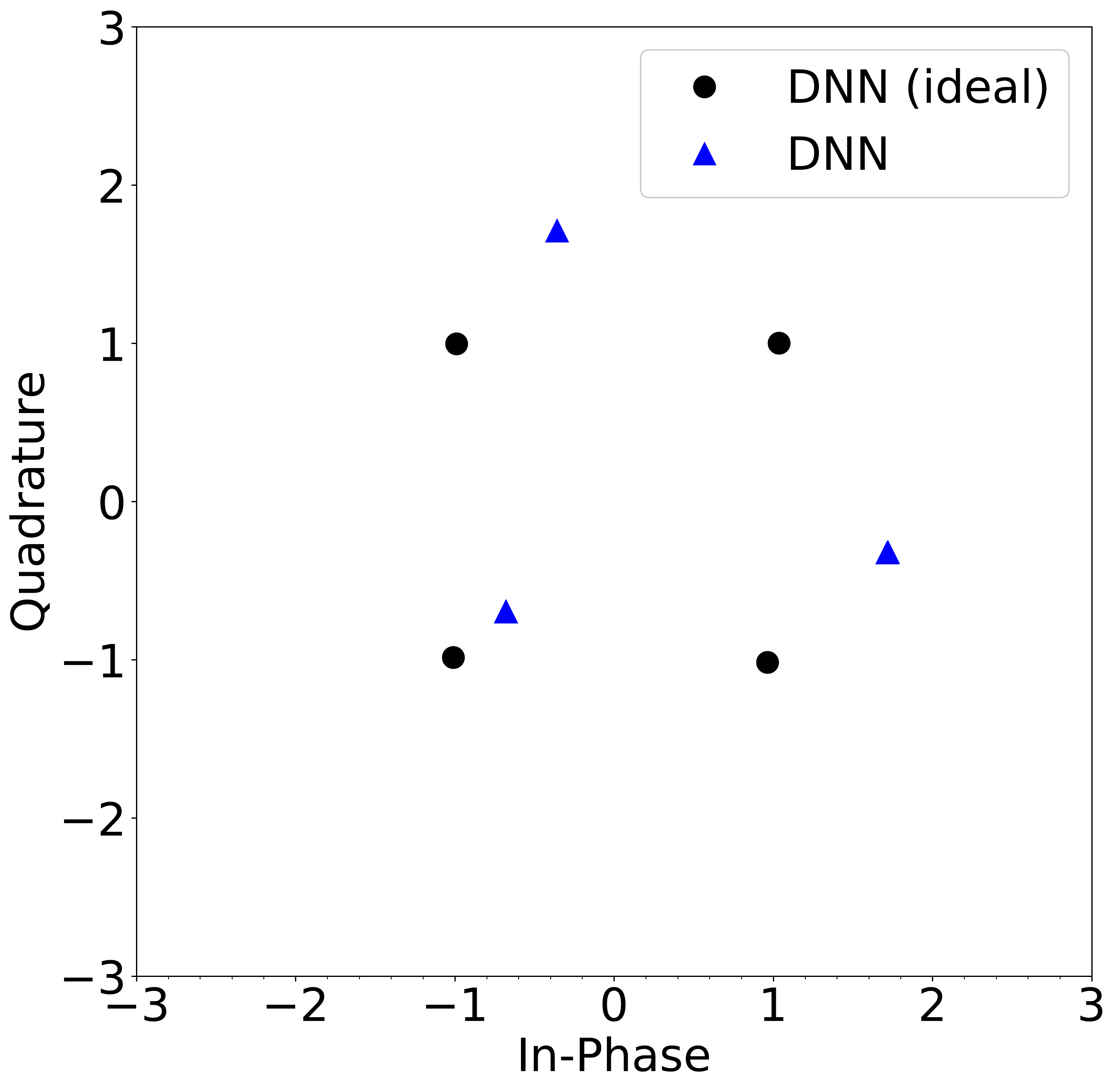} \label{4train0} }
 \hfil
 \subfloat[Training SNR: $5$~dB]{ \includegraphics[width=0.45\hsize]{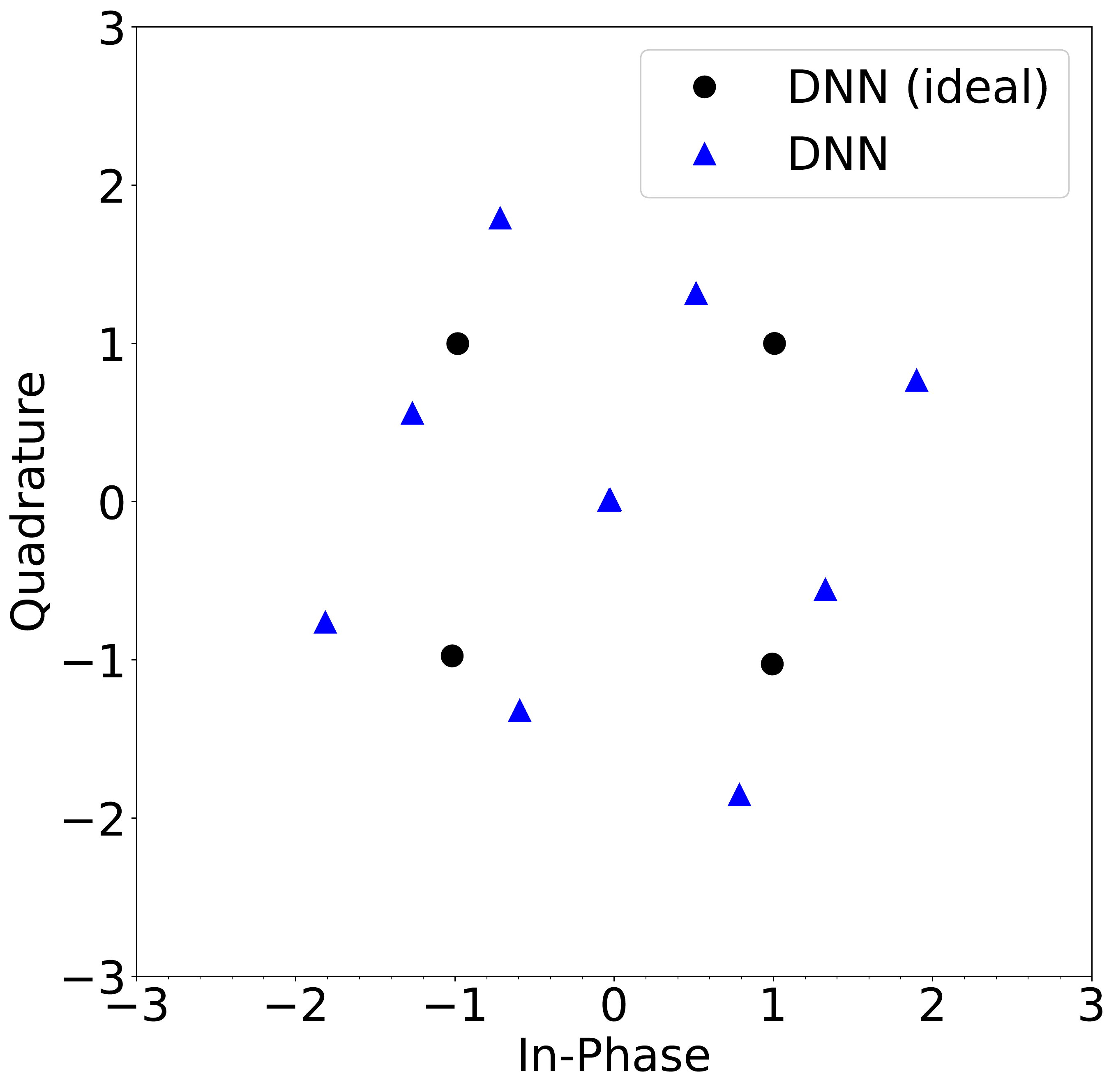} \label{4train5} }
\caption{Signal constellations transmitted from the relay to terminals A and B in the BC step for the proposed scheme. 4-QAM is used in the MA step.}
\label{fig:qpsk_conste}
\end{figure}

\subsection{AWGN Channel}
In this subsection, we evaluate the performance of the proposed scheme over the AWGN channel in terms of the achievable sum rate.
We use rectified linear unit (ReLU) activations in the DNN architecture.
As a benchmark, we also consider an ideal case where $S_A, S_B$ and $X_A, X_B$ are available at the relay DNN $\mathcal{M}_R$.
Although this ideal DNN scenario may not be practical, it is useful for seeing an upper-bound on the room for performance improvement
by enhancing the detection of the information $S_A, S_B$ and $X_A, X_B$ at the relay.

\subsubsection{4-QAM}
Fig.~\ref{fig:qpsk_awgn} shows the performance comparison of the proposed DNN-based scheme
and the conventional AF relaying over AWGN channel in terms of the achievable sum rate defined as \eqref{eq:thp}.
Here, we consider two bits per symbol constellation, i.e., 4-QAM.
From this figure, it is observed that the proposed scheme outperforms the conventional AF relaying by more than $1$\,dB.
In particular, the proposed scheme offers the significant performance gain over the AF relaying systems in the low SNR region,
where the improvement of $1$ bps/Hz is observed in the achievable sum rate at an SNR of 0\,dB.
Also we can see from this figure that the proposed scheme has a large performance gap from the ideal case, which is approximately $3$\,dB.
This indicates that the performance of proposed scheme has a potential improvement
if the relay detects the information $S_A, S_B$ and $X_A, X_B$ more accurately.

Fig.~\ref{fig:qpsk_conste} shows the signal constellation that the relay broadcasts to terminals A and B in the BC step.
In this figure, the Gaussian noise at the relay $Z_R$ is not shown.
From this figure, it is observed that when the relay knows the information $S_A, X_A$ and $S_B, X_B$ (ideal case),
the relay broadcasts a 4-QAM constellation to terminals in all cases, which is equivalent to XOR denoising \cite{koike2009optimized}.
Meanwhile, when the relay has no such information and the channel SNR is low, it sends $3$-ary constellation.
This indicates that the proposed DNN automatically controls the transmission rate, depending on the channel condition.
For the training SNR of $5$--$20$\,dB, the relay sends $9$-QAM, which is similar to the conventional AF scheme  (since the result for the training SNR $10$--$20$\,dB is similar to that of $5$\,dB, they were omitted).

\begin{figure}[t]
\centering
\includegraphics[width=0.9\hsize]{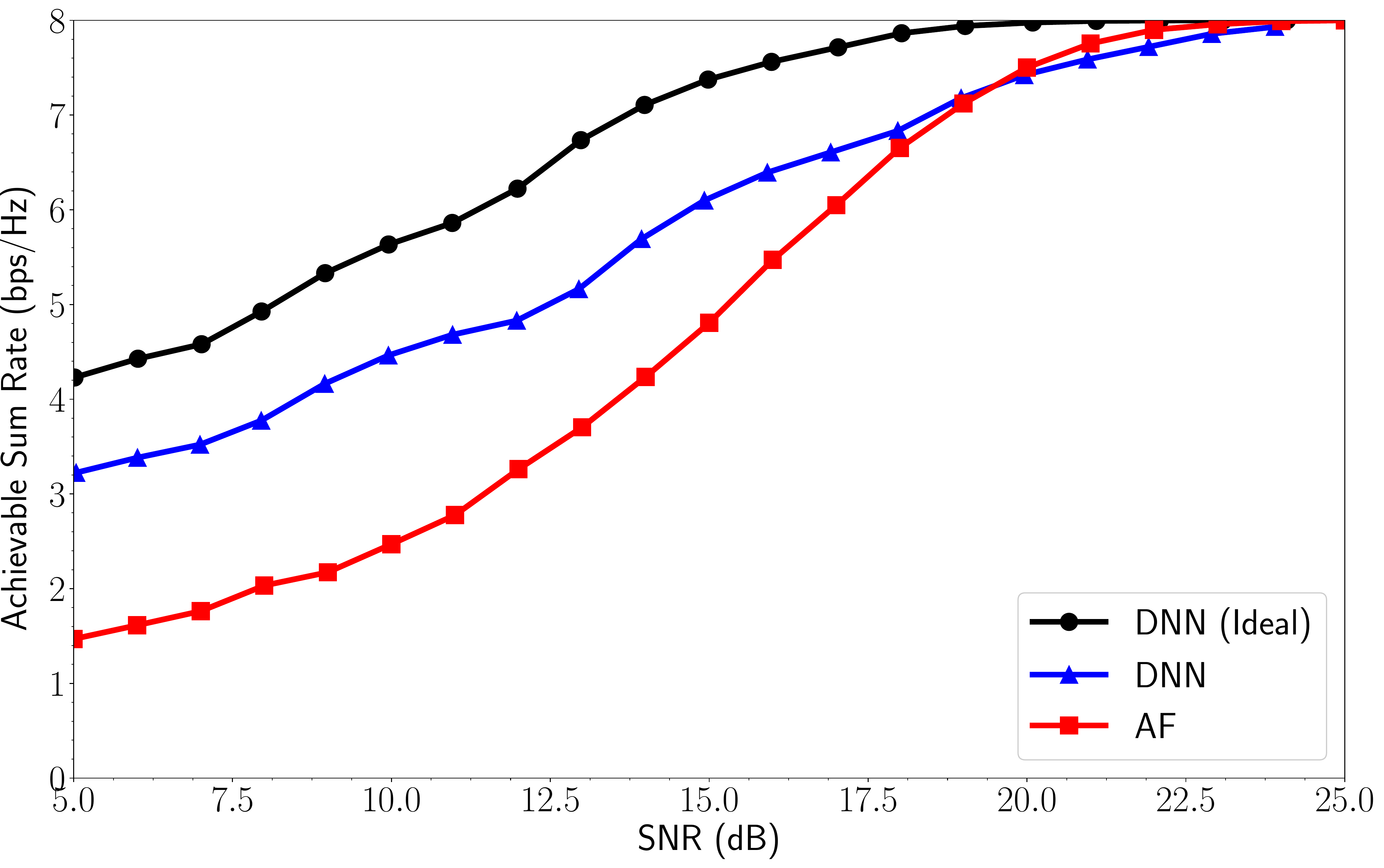}
\caption{Achievable sum rates of the proposed DNN relaying and the conventional AF relaying over AWGN channel with 16-QAM.}
\label{fig:16qam_awgn}
\end{figure}
\begin{figure}[t]
 \centering
 \subfloat[Training SNR: $5$~dB]{ \includegraphics[width=0.45\hsize]{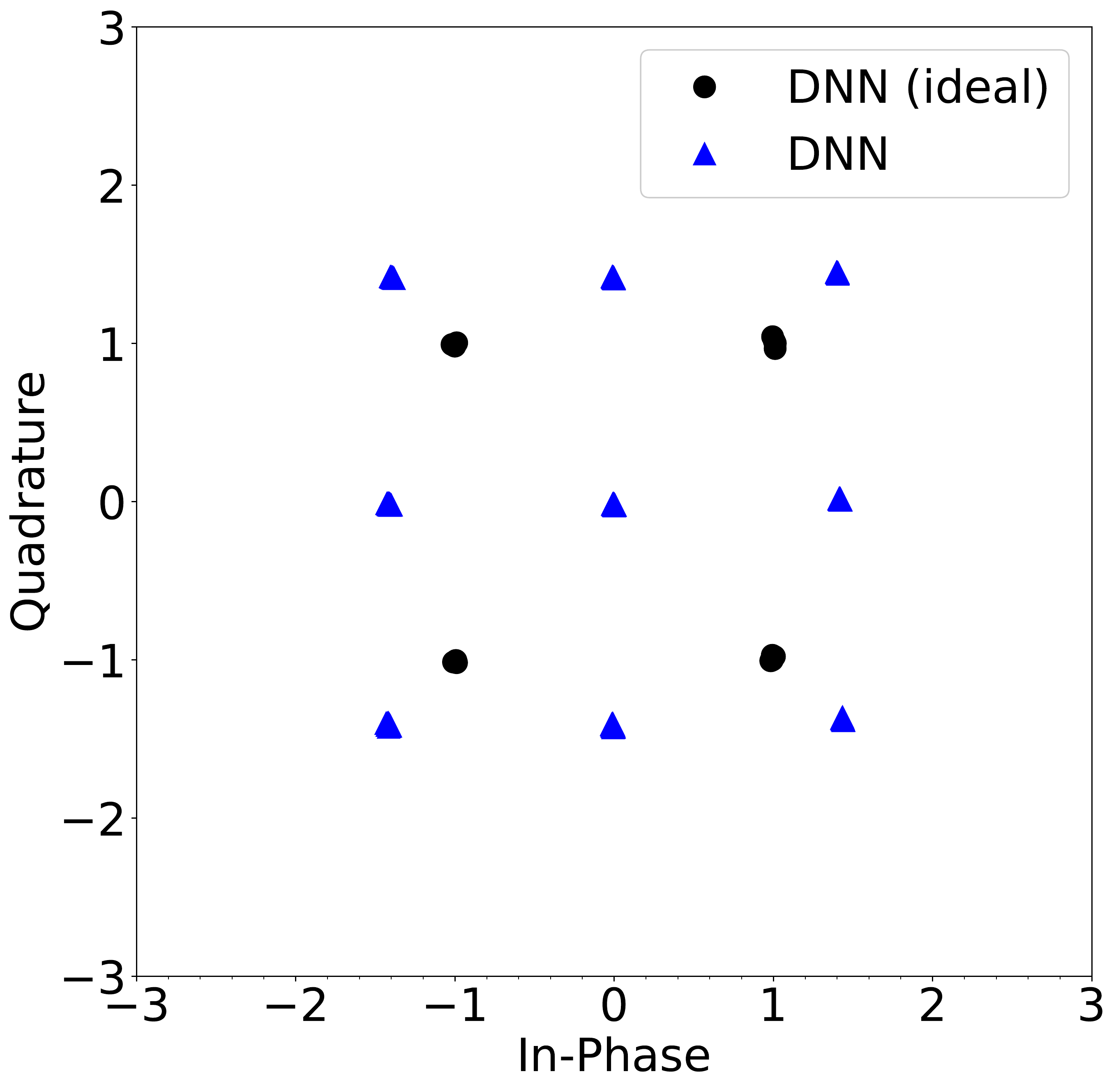} \label{16train5} }
 \hfil
 \subfloat[Training SNR: $10$~dB]{ \includegraphics[width=0.45\hsize]{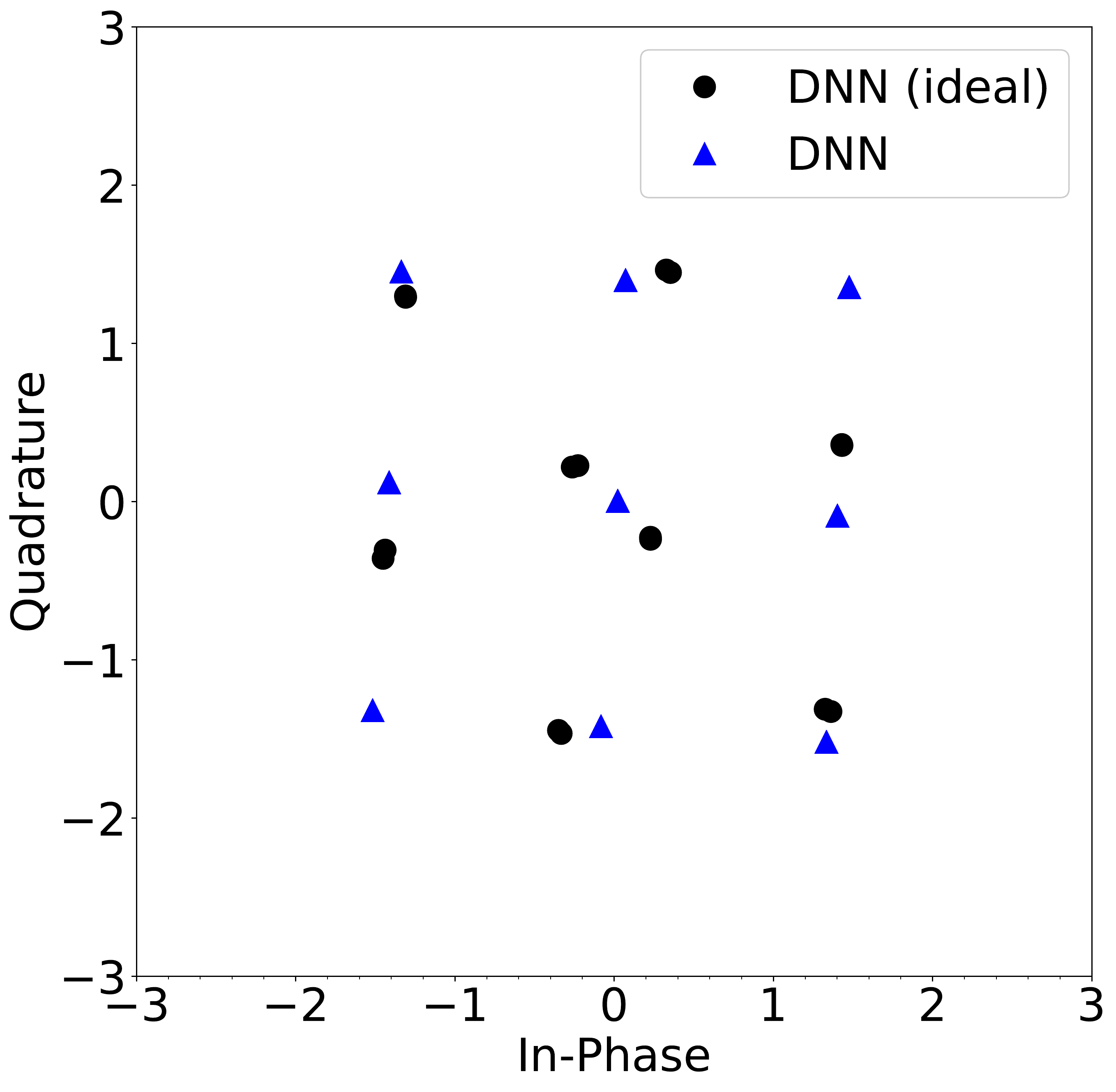} \label{16train10} }
 \hfil \\
 \subfloat[Training SNR: $15$~dB]{ \includegraphics[width=0.45\hsize]{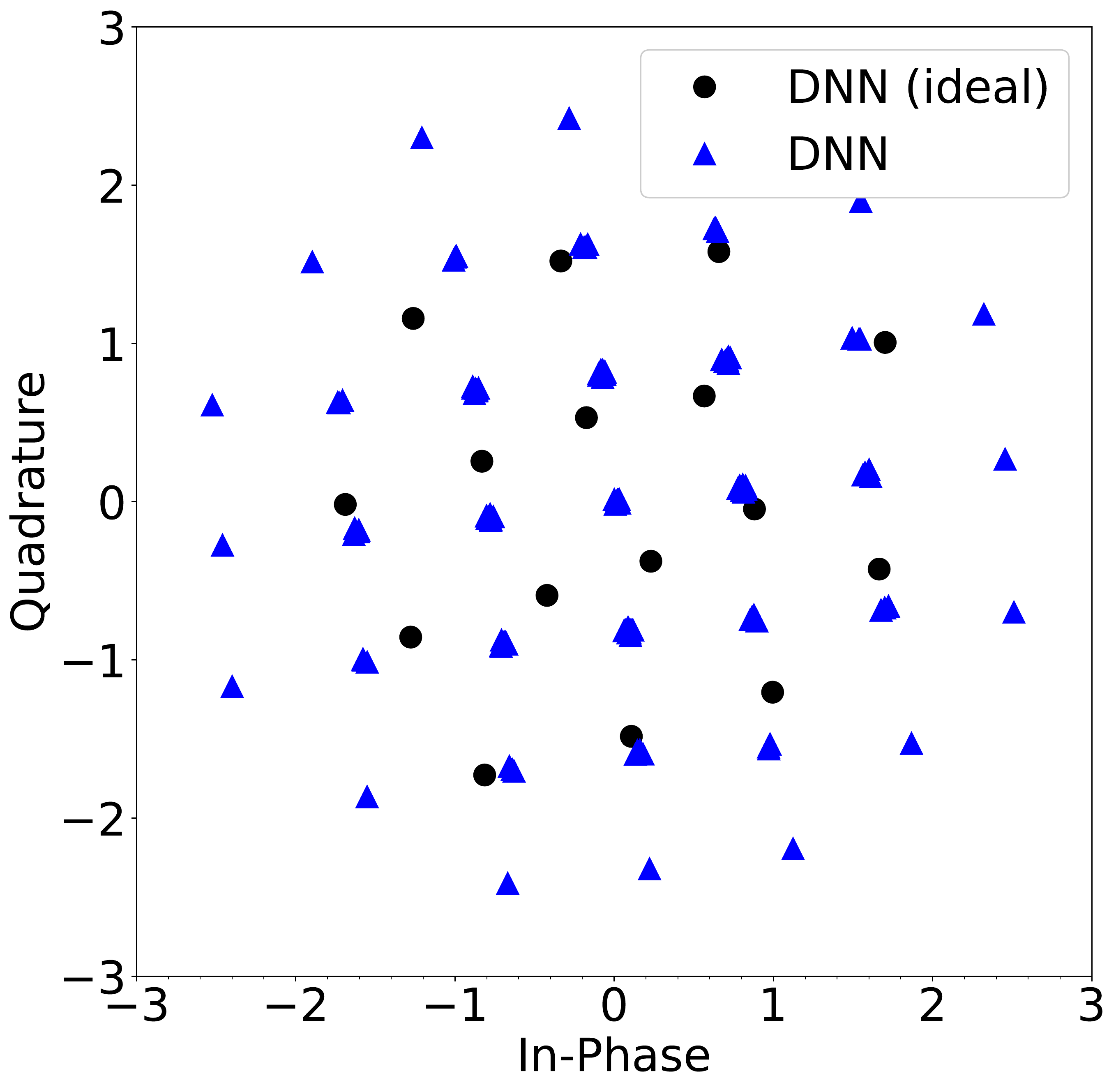} \label{16train15} }
 \hfil
 \subfloat[Training SNR: $25$~dB]{ \includegraphics[width=0.45\hsize]{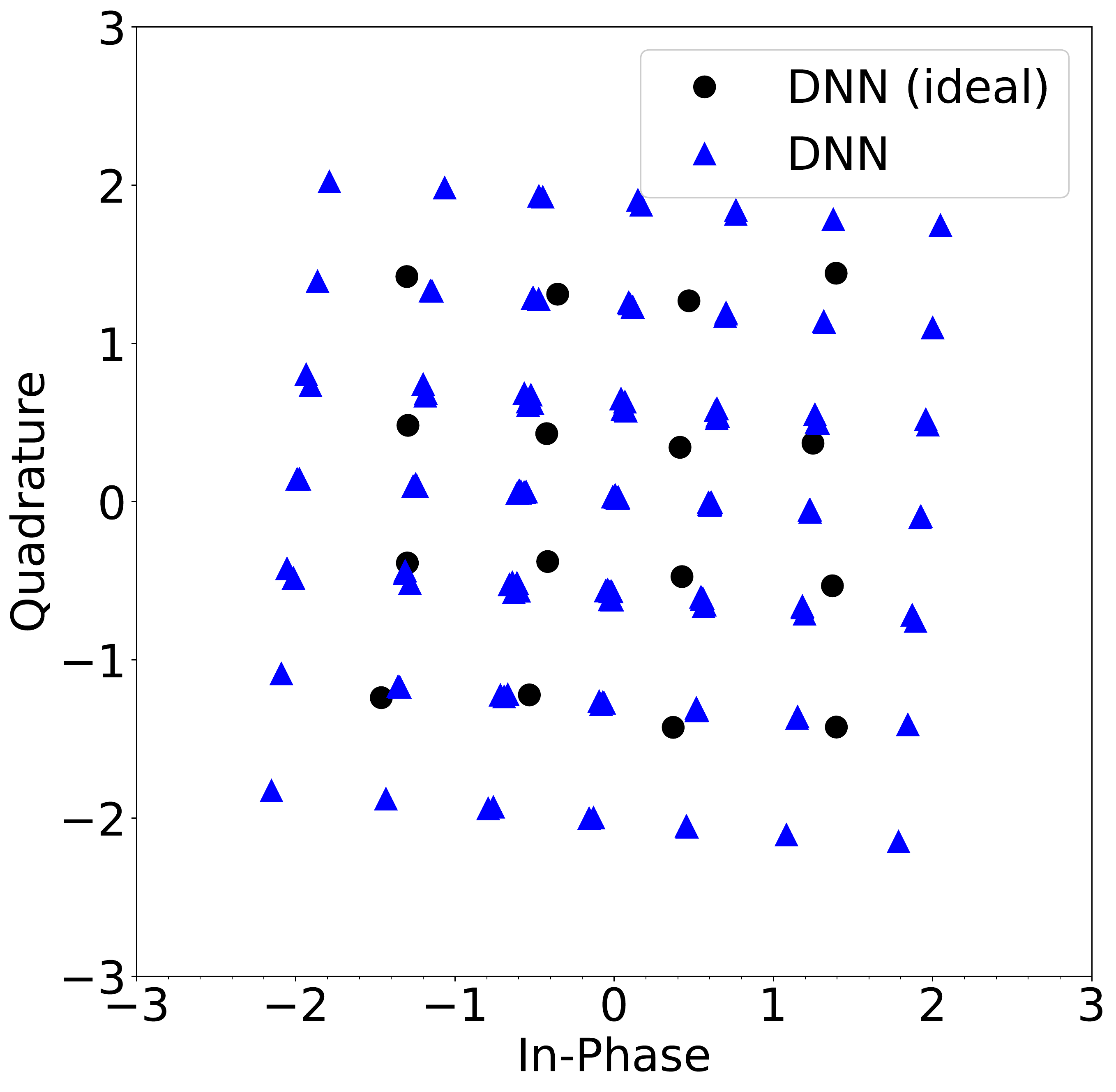} \label{16train25} }
\caption{Signal constellations transmitted from the relay to terminals A and B in the BC step for the proposed scheme. 16-QAM is used in the MA step.}
\label{fig:16qam_conste}
\end{figure}

\subsubsection{16-QAM}
Fig.~\ref{fig:16qam_awgn} shows the performance with 16-QAM signaling over the AWGN channel.
We observed from this figure that also in this case, the proposed scheme significantly outperforms the AF relaying scheme in the low SNR region,
since it flexibly controls the transmission rate depending on the channel condition.
On the other hands,  in the high SNR region, proposed DNN is outperformed by the conventional AF scheme.
This is because the AF performs the optimal ML detection at terminals A and B,
while in our proposed scheme, the DNN demodulators try to approximate ML detection with a nonlinear function.
Also, the proposed scheme still has a large performance gap from the ideal case.

In Fig.~\ref{fig:16qam_conste} we show the signal constellation that the relay sends to terminals A and B in the BC step.
For the purpose of plotting the DNN modulator $\mathcal{M}_R$, the Gaussian noise $Z_R$ is again not shown.
For the training SNR of $5$\,dB, the DNN compresses the constellation to $4$-QAM in the ideal case, 
while the relay broadcasts $9$-QAM to terminals A and B when it has no information about $S_A, X_A$ and $S_B, X_B$.
The DNN without the information still sends $9$-QAM when the training SNR is $5$\,dB,
whereas the ideal case transmits $8$-ary constellation.
For relatively higher training SNRs, e.g., $15$\,dB, constellations become circular for both cases, 
while the ideal case compresses the constellation more effectively (similar results are observed for training SNR of $20$\,dB). 
For even higher training SNRs, the ideal case sends $16$-QAM, which is equivalent to XOR denoising \cite{koike2009optimized}.
On the other hands, the DNN without the information sends almost $49$-QAM constellation, which is equivalent to the AF scheme.

\subsection{Frequency-Flat Rayleigh Fading Channel}

\begin{figure}[t]
\centering
\includegraphics[width=0.9\hsize]{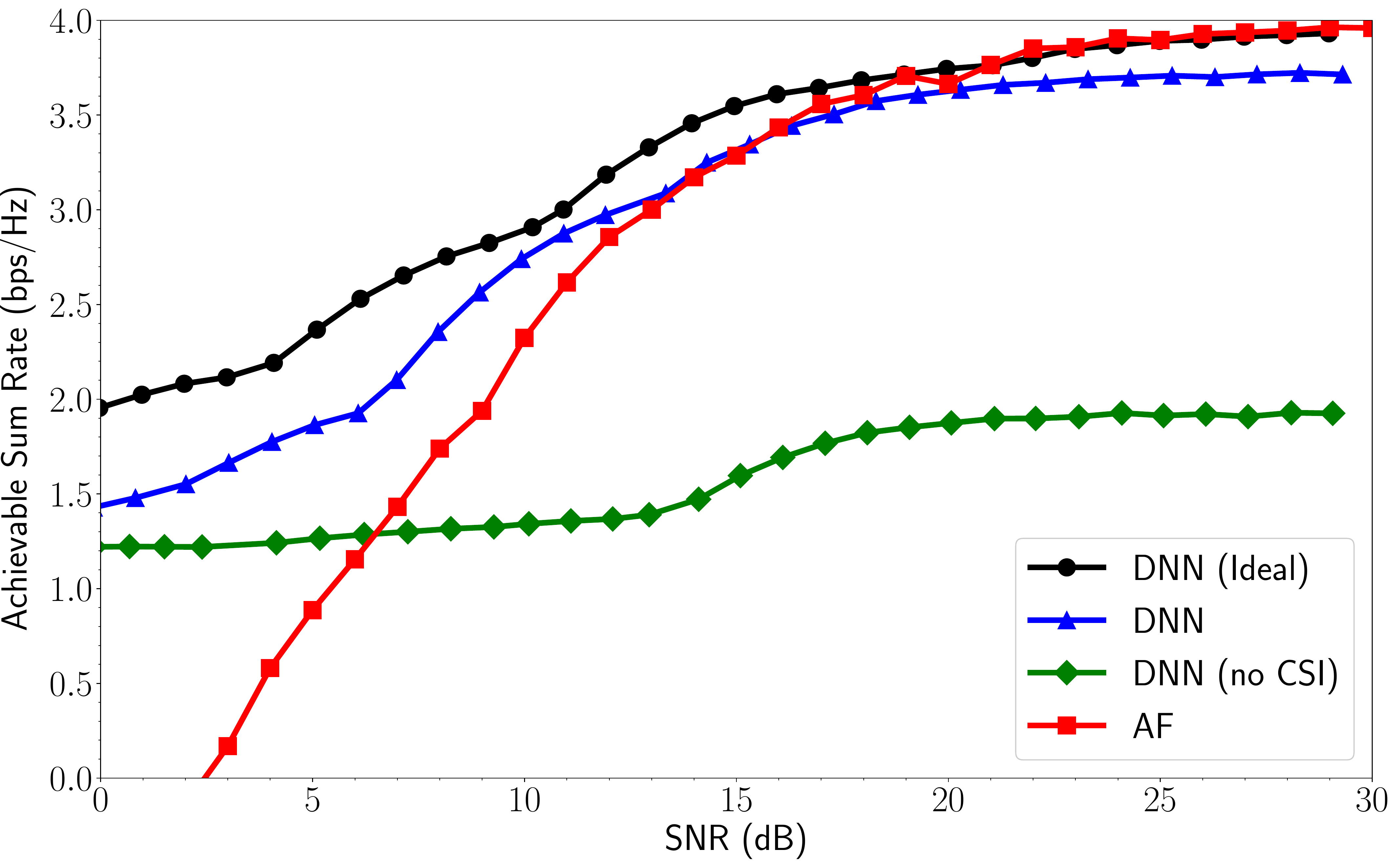}
\caption{Achievable sum rates of the proposed DNN relaying and the conventional AF relaying over Rayleigh fading channel with 4-QAM.}
\label{fig:qpsk_fading}
\end{figure}
In what follows, we show the simulation results for $4$-QAM signaling over frequency-flat block Rayleigh fading channels, where channel coefficients are identical in a mini-batch.
We use $\mathrm{tanh}$ activations in this subsection.
We consider the following three scenarios in the following:
\begin{itemize}
\item No CSI is available: denoted by ``DNN (no CSI).''
\item Perfect CSI is available at both terminals and the relay. Additionally, $S_A, S_B$ and $X_A, X_B$ are available at the relay: denoted by ``DNN (Ideal).''
\item Perfect CSI is available at both terminals and the relay: denoted by ``DNN.''
\end{itemize}

The achievable sum rate for frequency-flat Rayleigh fading with 4-QAM is shown in Fig.~\ref{fig:qpsk_fading}.
In this figure, we can see a significant performance gain greater than $2$\,dB by the proposed scheme over the AF scheme in the low SNR region.
Also, we observe that even in the high SNR region, e.g., $30$\,dB, the proposed scheme still has a gap from the ideal case,
which indicates that the proposed scheme has a room to improve the performance by improving the detection at the relay.
It is also observed that the achievable sum rate of the proposed scheme with no CSI case does not increase more than about the half of the best achievable sum rate for this setting. 
This is because the amplitude and phase information in signal constellation can be useless for unknown CSI, and an on-off keying type constellation might have been learned.


\section{Conclusion}
\label{sec:con}
In this paper, we proposed a new application of deep learning to PNC in wireless relay networks,
where the signal constellation mapper/demapper is performed by DNNs.
The DNN is trained such that the GMI is directly maximized,
which is an appropriate performance measure for practical wireless systems with soft-decision channel decoders.
Simulation results demonstrated that the proposed DNN-based PNC scheme offers
a significant performance gain over the conventional AF relaying systems 
to achieve higher sum rates.

Since it was observed that there may exist a potential room for the significant performance improvement
by enhancing the detection at the relay,
multi-task learning \cite{caruana1997multitask} for taking both the end-to-end cross entropy and the detection error at the relay into account may be helpful.
Also, since our work in this paper is limited to the design of the constellation mapper/demapper,
our future work may include the joint design of channel coding for the DNN-based relaying network.


\bibliographystyle{IEEEtran}
\bibliography{IEEEabrv,matsumine}

\end{document}